\documentclass[useAMS,usenatbib]{mn2e}
\usepackage{amsmath,amsfonts,latexsym,graphicx,amssymb,times,appendix}
\usepackage{color}

\title[Investigation into MOPED]{An investigation into the Multiple Optimised Parameter Estimation and Data compression algorithm}
\author[Graff, Hobson, and Lasenby]{Philip Graff$^1$, Michael P. Hobson$^1$, and Anthony Lasenby$^{1,2}$\\
		$^1$Astrophysics Group, Cavendish Laboratory, JJ Thomson Avenue, Cambridge CB3 0HE, UK\\
		$^2$Kavli Institute for Cosmology, Madingley Road, Cambridge CB3 0HA, UK}
\date{\today}

\pagerange{\pageref{firstpage}--\pageref{lastpage}} \pubyear{2010}

\begin{document}

\label{firstpage}

\maketitle

\begin{abstract}
We investigate the use of the Multiple Optimised Parameter Estimation and Data compression algorithm (MOPED) for data compression and faster evaluation of likelihood functions.  Since MOPED only guarantees maintaining the Fisher matrix of the likelihood at a chosen point, multimodal and some degenerate distributions will present a problem.  We present examples of scenarios in which MOPED does faithfully represent the true likelihood but also cases in which it does not.  Through these examples, we aim to define a set of criteria for which MOPED will accurately represent the likelihood and hence may be used to obtain a significant reduction in the time needed to calculate it.  These criteria may involve the evaluation of the full likelihood function for comparison.
\end{abstract}


\section{Introduction}
Multiple Optimised Parameter Estimation and Data compression (MOPED;~\cite{MOPED1}) is a patented algorithm for the compression of data and the speeding up of the evaluation of likelihood functions in astronomical data analysis and beyond. It becomes particularly useful when the noise covariance matrix is dependent upon the parameters of the model and so must be calculated and inverted at each likelihood evaluation. However, such benefits come with limitations. Since MOPED only guarantees maintaining the Fisher matrix of the likelihood at a chosen point, multimodal and some degenerate distributions will present a problem. In this paper we report on some of the limitations of the application of the MOPED algorithm. In the cases where MOPED does accurately represent the likelihood function, however, its compression of the data and consequent much faster likelihood evaluation does provide orders of magnitude improvement in runtime.  In~\cite{MOPED1}, the authors demonstrate the method by analysing the spectra of galaxies and in~\cite{MOPED2} they illustrate the benefits of MOPED for estimation of the CMB power spectrum. The problem of ``badly'' behaved likelihoods was found by~\cite{Protopapas} for the problem of light transit analysis; nonetheless, the authors present a solution that still allows MOPED to provide a large speed increase.

We begin by introducing MOPED in Section~2 and define the original and MOPED likelihood functions, along with comments on the potential speed benefits of MOPED.  In Section~3 we introduce an astrophysical scenario where we found that MOPED did not accurately portray the true likelihood function. In Section~4 we expand upon this scenario to another where MOPED is found to work and to two other scenarios where it does not. We present a discussion of the criteria under which we believe MOPED will accurately represent the likelihood in Section~5, as well as a discussion of an implementation of the solution provided by~\cite{Protopapas}.

\section{Data Compression with MOPED}
Full details of the MOPED method are given in~\cite{MOPED1}, here we will only present a limited introduction.

We begin by defining our data as a vector, ${\bf x}$. Our model describes ${\bf x}$ by a signal plus random noise,
\begin{equation}
\label{eq:defdata}
{\bf x} = {\bf u}(\btheta_T) + {\bf n}(\btheta_T),
\end{equation}
where the signal is given by a vector ${\bf u}(\btheta)$ that is a function of the set of parameters $\btheta=\{\theta_i\}$ defining our model, and the true parameters are given by $\btheta_T$. The noise is assumed to be Gaussian with zero mean and noise covariance matrix $\mathcal{N}_{jk}=\left<n_j n_k\right>$, where the angle brackets indicate an ensemble average over noise realisations (in general this matrix may also be a function of the parameters $\btheta$). The full likelihood for $N$ data points in ${\bf x}$ is given by
\begin{eqnarray}
\label{eq:full_like}
\mathcal{L}_{\textrm{Original}}(\btheta) &=& \frac{1}{(2\pi)^{N/2} \sqrt{\left|\mathcal{N}(\btheta)\right|}} \times \notag \\
&&\exp{\left\{-\frac{1}{2}[{\bf x}-{\bf u}(\btheta)]^{\textrm{T}} \mathcal{N}(\btheta)^{-1} [{\bf x}-{\bf u}(\btheta)]\right\}}.
\end{eqnarray}
At each point, then, this requires the calculation of the determinant and inverse of an $N \times N$ matrix. Both scale as $N^3$, so even for smaller datasets this can become cumbersome.

MOPED allows one to eliminate the need for this matrix inversion by compressing the $N$ data points in ${\bf x}$ into $M$ data values, one for each parameters of the model. Additionally, MOPED creates the compressed data values such that they are independent and have unit variance, further simplifying the likelihood calculation on them to an $O(M)$ operation. Typically, $M \ll N$ so this gives us a significant increase in speed. A single compression is done on the data, ${\bf x}$, and then again for each point in parameter space where we wish to compute the likelihood. The compression is done by generating a set of weighting vectors, ${\bf b}_i(\btheta_F)$  ($i=1 \ldots M$), from which we can generate a set of MOPED components from the theoretical model and data,
\begin{equation}
\label{eq:genMcomp}
y_i(\btheta_F) \equiv {\bf b}_i(\btheta_F) \cdot {\bf x} = {\bf b}_i(\btheta_F)^{\textrm{T}} {\bf x}.
\end{equation}
Note that the weighting vectors must be computed at some assumed fiducial set of parameter values, $\btheta_F$.  The only choice that will truly maintain the likelihood peak is when the fiducial parameters are the true parameters, but obviously we will not know these in advance for real analysis situations.  Thus, we can choose our fiducial model to be anywhere and iterate the procedure, taking our likelihood peak in one iteration as the fiducial model for the next iteration.  This process will converge very quickly, and may not even be necessary in some instances.  For our later examples, since we do know the true parameters we will use these as the fiducial ($\btheta_F = \btheta_T$) in order to remove this as a source of confusion (all equations, however, are written for the more general case).  Note that the true parameters, $\btheta_T$, will not necessarily coincide with the peak $\hat{\btheta}_O$ of the original likelihood or the peak $\hat{\btheta}_M$ of the MOPED likelihood (see below).

The weighting vectors must be generated in some order so that each subsequent vector (after the first) can be made orthogonal to all previous ones. We begin by writing the derivative of the model with respect to the $i$th parameter as $\tfrac{\partial{\bf u}}{\partial\theta_i}|_{\btheta_F}={\bf u}_{,i}(\btheta_F)$. This gives us a solution for the first weighting vector, properly normalised, of
\begin{equation}
\label{eq:b1eqn}
{\bf b}_1(\btheta_F) = \frac{\mathcal{N}(\btheta_F)^{-1}{\bf u}_{,1}(\btheta_F)}{\sqrt{{\bf u}_{,1}(\btheta_F)^{\textrm{T}}\mathcal{N}(\btheta_F)^{-1}{\bf u}_{,1}(\btheta_F)}}.
\end{equation}
The first compressed value is $y_1(\btheta_F)={\bf b}_1(\btheta_F)^{\textrm{T}}{\bf x}$ and will weight up the data combination most sensitive to the first parameter. The subsequent weighting vectors are made orthogonal by subtracting out parts that are parallel to previous vectors, and are normalized. The resulting formula for the remaining weighting vectors is
\begin{eqnarray}
&&{\bf b}_m(\btheta_F) =
\end{eqnarray}
\begin{equation*}
\label{eq:bmeqn}
\frac{\mathcal{N}(\btheta_F)^{-1}{\bf u}_{,m}(\btheta_F)-\sum_{q=1}^{m-1}{({\bf u}_{,m}(\btheta_F)^{\textrm{T}}{\bf b}_q(\btheta_F)){\bf b}_q(\btheta_F)}}{\sqrt{{\bf u}_{,m}(\btheta_F)^{\textrm{T}}\mathcal{N}(\btheta_F)^{-1}{\bf u}_{,m}(\btheta_F)-\sum_{q=1}^{m-1}{({\bf u}_{,m}(\btheta_F)^{\textrm{T}}{\bf b}_q(\btheta_F))^2}}},
\end{equation*}
where $m=2 \ldots M$. Weighting vectors generated with Equations~\eqref{eq:b1eqn} and~\eqref{eq:bmeqn} form an orthnomal set with respect to the noise covariance matrix so that
\begin{equation}
\label{eq:borthonorm}
{\bf b}_i({\bf \theta}_F)^{\textrm{T}} \mathcal{N}({\bf \theta}_F) {\bf b}_j({\bf \theta}_F) = \delta_{ij}.
\end{equation}
This means that the noise covariance matrix of the compressed values $y_i$ is the identity, which significantly simplifies the likelihood calculation. The new likelihood function is given by 
\begin{eqnarray}
\label{eq:newlike}
\mathcal{L}_{\textrm{MOPED}}({\bf \theta}) &=& \frac{1}{(2\pi)^{M/2}} \times \notag \\
&&\exp{\left\{ -\frac{1}{2} \sum_{i=1}^{M}{(y_i({\bf \theta}_F)-\left<y_i\right>({\bf \theta};{\bf \theta}_F))^2} \right\}},
\end{eqnarray}
where $y_i(\btheta_F)={\bf b}_i(\btheta_F)^{\textrm{T}}{\bf x}$ represents the compressed data and $\left<y_i\right>(\btheta;\btheta_F) = {\bf b}_i(\btheta_F)^{\textrm{T}} {\bf u}(\btheta)$ represents the compressed signal. This is a much easier likelihood to calculate and is time-limited by the generation of a new signal template instead of the inversion of the noise covariance matrix.  The peak value of the MOPED likelihood function is not guaranteed to be unique as there may be other points in the original parameter space that map to the same point in the compressed parameter space; this is a characteristic that we will investigate.

MOPED implicity assumes that the covariance matrix, $\mathcal{N}$, is independent of the parameters. With this assumption, a full likelihood calculation with $N$ data points would require only an $O(N^2)$ operation at each point in parameter space (or $O(N)$ if $\mathcal{N}$ is diagonal). In MOPED, however, the compression of the theoretical data is an $O(MN)$ linear operation followed by an $O(M)$ likelihood calculation. Thus, one loses on speed if $\mathcal{N}$ is diagonal but gains by a factor of $N/M$ otherwise. For the data sets we will analyze, $N/M > 100$. We begin, though, by assuming a diagonal $\mathcal{N}$ for simplicity, recognizing that this will cause a speed reduction but that it is a necessary step before addressing a more complex noise model. One can iterate the parameter estimation procedure if necessary, taking the maximum likelihood point found as the new fiducial and re-analyzing (perhaps with tighter prior constraints); this procedure is recommended for MOPED in~\cite{MOPED1}, but is not always found to be necessary. MOPED has the additional benefit that the weighting vectors, ${\bf b}_i$, need only to be computed once provided the fiducial model parameters are kept constant over the analysis of different data sets. Computed compressed parameters, $\left<y_i\right>$, can also be stored for re-use and require less memory than storing the entire theoretical data set.

\section{Simple Example With One Parameter}
In order to demonstrate some of the limitations of the applicability of the MOPED algorithm, we will consider a few test cases.  These originate in the context of gravitational wave data analysis for the \emph{Laser Interferometer Space Antenna} (\emph{LISA}) since it is in this scenario that we first discovered such cases of failure.  The full problem is seven-dimensional parameter estimation, but we have fixed most of these variables to their known true values in the simulated data set in order to create a lower-dimensional problem that is simpler to analyse.

We consider the case of a sine-Gaussian burst signal present in the LISA detector.  The short duration of the burst with respect to the motion of LISA allows us to use the static approximation to the response.  In frequency space, the waveform is described by~(\cite{Strings})
\begin{equation}
\label{eq:sineGuass}
\tilde{h}(f)=A\tfrac{Q}{f} \exp{\left\{-\tfrac{1}{2}Q^2(\tfrac{f-f_c}{f_c})^2\right\}}\exp(2\pi \imath t_0 f).
\end{equation}
Here $A$ is the dimensionless amplitude factor; $Q$ is the width of the Gaussian envelope of the burst measured in cycles; $f_c$ is the central frequency of the oscillation being modulated by the Gaussian envelope; and $t_0$ is the central time of arrival of the burst.  This waveform is further modulated by the sky position of the burst source, $\theta$ and $\phi$, and the burst polarisation, $\psi$, as they project onto the detector.  The one-sided noise power spectral density of the LISA detector is given by~(\cite{Strings})
\begin{eqnarray}
S_h(f) &=& 16 \sin^2(2\pi f t_L) \times \nonumber \\
&&\left(2\left(1+\cos(2\pi f t_L)+\cos^2(2\pi f t_L)\right)S_{\rm pm}(f) \right. \nonumber \\ 
&& \left.+ \left(1+\cos(2\pi f t_L)/2\right)S_{\rm sn}f^2\right),\label{eq:sha} \\
S_{\rm pm}(f) &=&  \left(1+\left(\frac{10^{-4}{\rm Hz}}{f}\right)^2\right)\frac{S_{\rm acc}}{f^2},
\label{eq:specdens}
\end{eqnarray}
where $t_L=16.678$s is the light travel time along one arm of the LISA constellation, $S_{\rm acc}=2.5\times10^{-48}$Hz$^{-1}$ is the proof mass acceleration noise and $S_{\rm sn}=1.8\times10^{-37}$Hz$^{-1}$ is the shot noise.  This is independent of the signal parameters and all frequencies are independent of each other, so the noise covariance matrix is constant and diagonal.  This less computationally expensive example allows us to show some interesting examples.

\begin{figure}
\begin{center}
\includegraphics[width=3.25in]{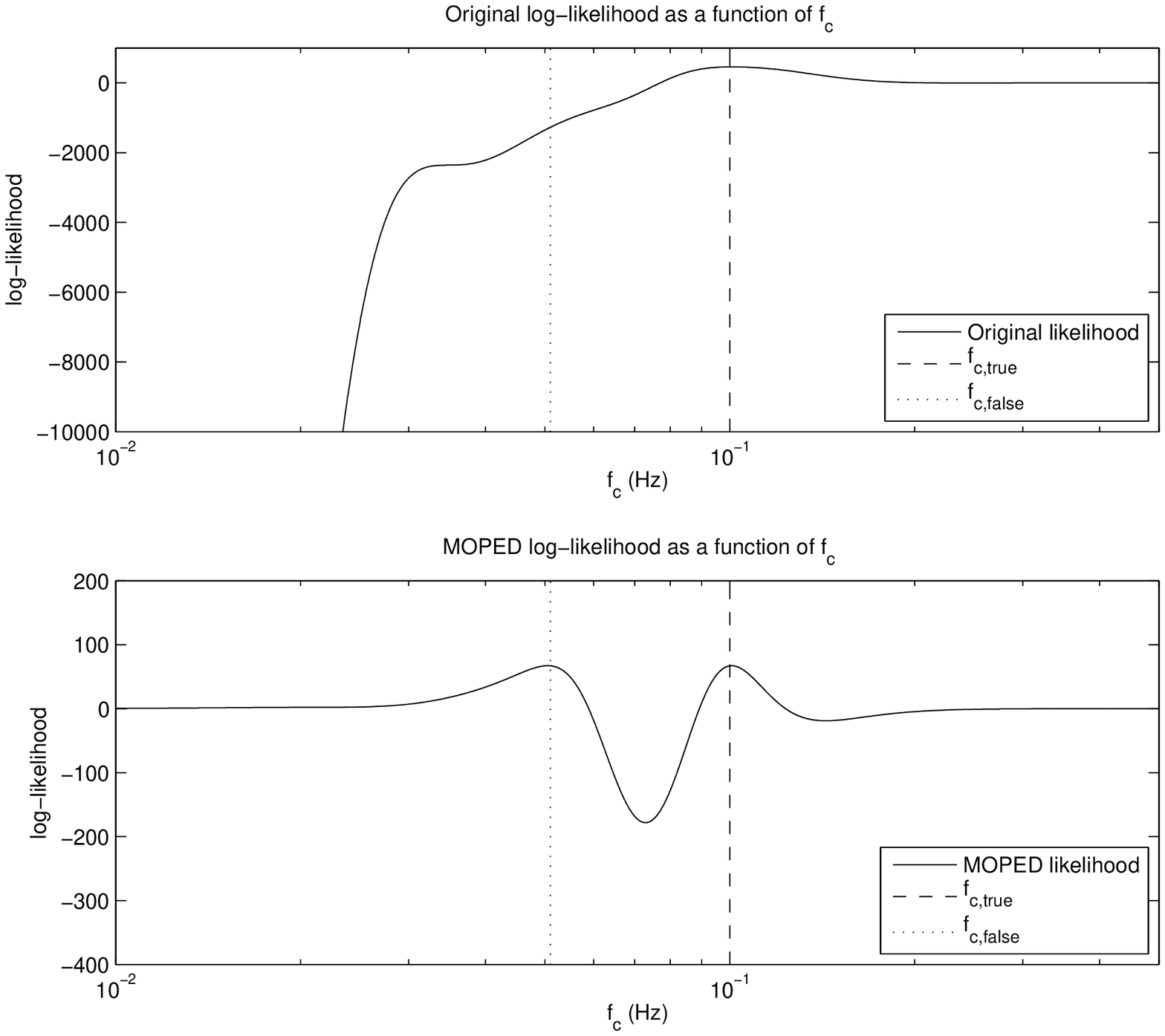}
\caption{The original and MOPED log-likelihoods as a function of $f_c$ for the chosen template.}
\label{fig:likecomp}
\end{center}
\end{figure}

We begin by taking the one-dimensional case where the only unknown parameter of the model is the central frequency of the oscillation, $f_c$.  We set $Q=5$ and $t_0=10^5$s; we then analyze a $2048$s segment of LISA data, beginning at $t=9.9\times 10^4$s, sampled at a $1$s cadence.  For this example, the data was generated with random noise (following the LISA noise power spectrum) at an SNR of $\sim 34$ with $f_{c,T}=0.1$Hz (thus $f_{c,F}=0.1$Hz for MOPED).  The prior range on the central frequency goes from $10^{-3}$Hz to $0.5$Hz.  $10,000$ samples uniformly spaced in $f_c$ were taken and their likelihoods calculated using both the original and MOPED likelihood functions.  The log-likelihoods are shown in Figure~\ref{fig:likecomp}.  Note that the absolute magnitudes are not important but the relative values within each plot are meaningful.  Both the original and MOPED likelihoods have a peak close to the input value $f_{c,T}$.

\begin{figure}
\begin{center}
\includegraphics[width=3.25in]{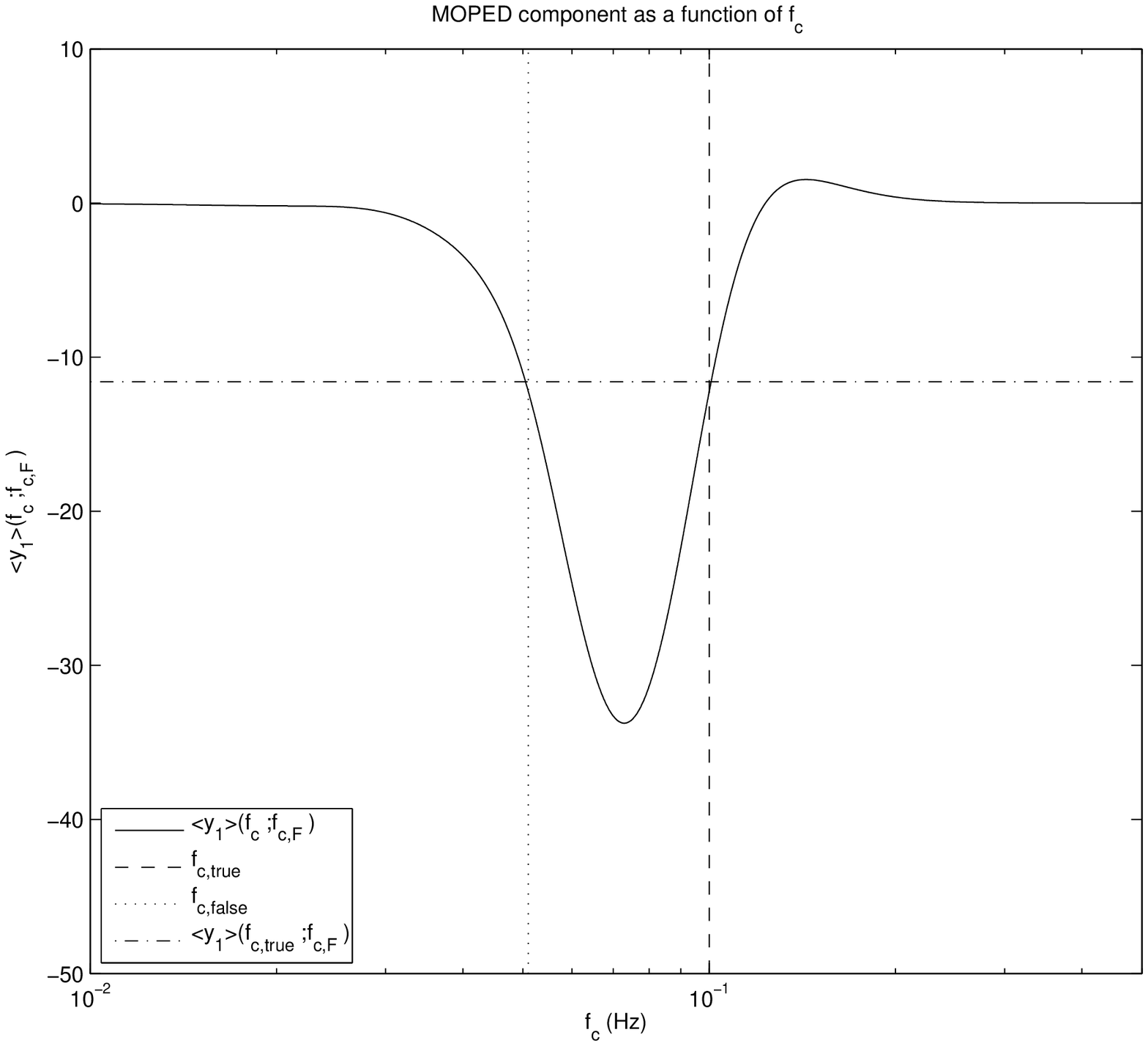}
\caption{The value of the MOPED compressed parameter as a function of the original frequency parameter.}
\label{fig:yf_vs_F}
\end{center}
\end{figure}

We see, however, that in going from the original to MOPED log-likelihoods, the latter also has a second peak of equal height at an incorrect $f_c$.  To see where this peak comes from, we look at the values of the compressed parameter $\left<y_1\right>(f_c;f_{c,F})$ as it varies with respect to $f_c$ as shown in Figure~\ref{fig:yf_vs_F}. The true compressed value peak occurs at $f_c \simeq 0.1$Hz, where $y_1(f_{c,F})=\left<y_1\right>(f_c;f_{c,F})$.  However, we see that there is another frequency that yields this exact same value of $\left<y_1\right>(f_c;f_{c,F})$; it is at this frequency that the second, incorrect peak occurs.
By creating a mapping from $f_c$ to $\left<y_1\right>(f_c;f_{c,F})$ that is not one-to-one, MOPED has created the possibility for a second solution that is indistinguishable in likelihood from the correct one.  This is a very serious problem for parameter estimation.

\section{Recovery in a 2 Parameter Case}
\begin{figure}
\begin{center}
\includegraphics[width=3.25in]{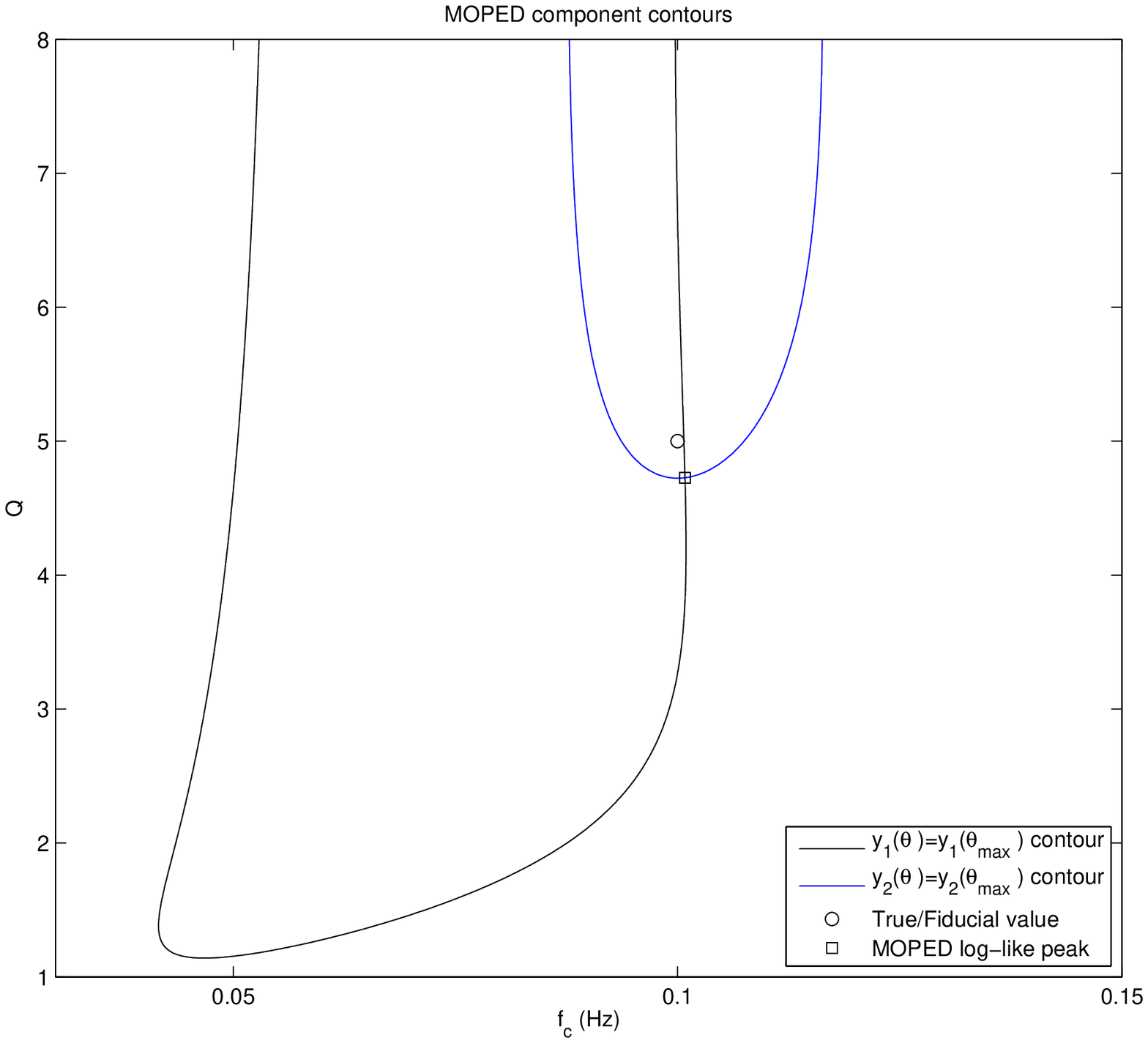}
\caption{Contours of $\left<y_1\right>(\btheta;\btheta_F) = \left<y_1\right>(\hat{\btheta}_M;\btheta_F)$ and $\left<y_2\right>(\btheta;\btheta_F) = \left<y_2\right>(\hat{\btheta}_M;\btheta_F)$ as they vary over $f_c$ and $Q$.  The one intersection is the true maximum likelihood peak.}
\label{fig:YtrueContours}
\end{center}
\end{figure}

\begin{figure}
\begin{center}
\includegraphics[width=3.25in]{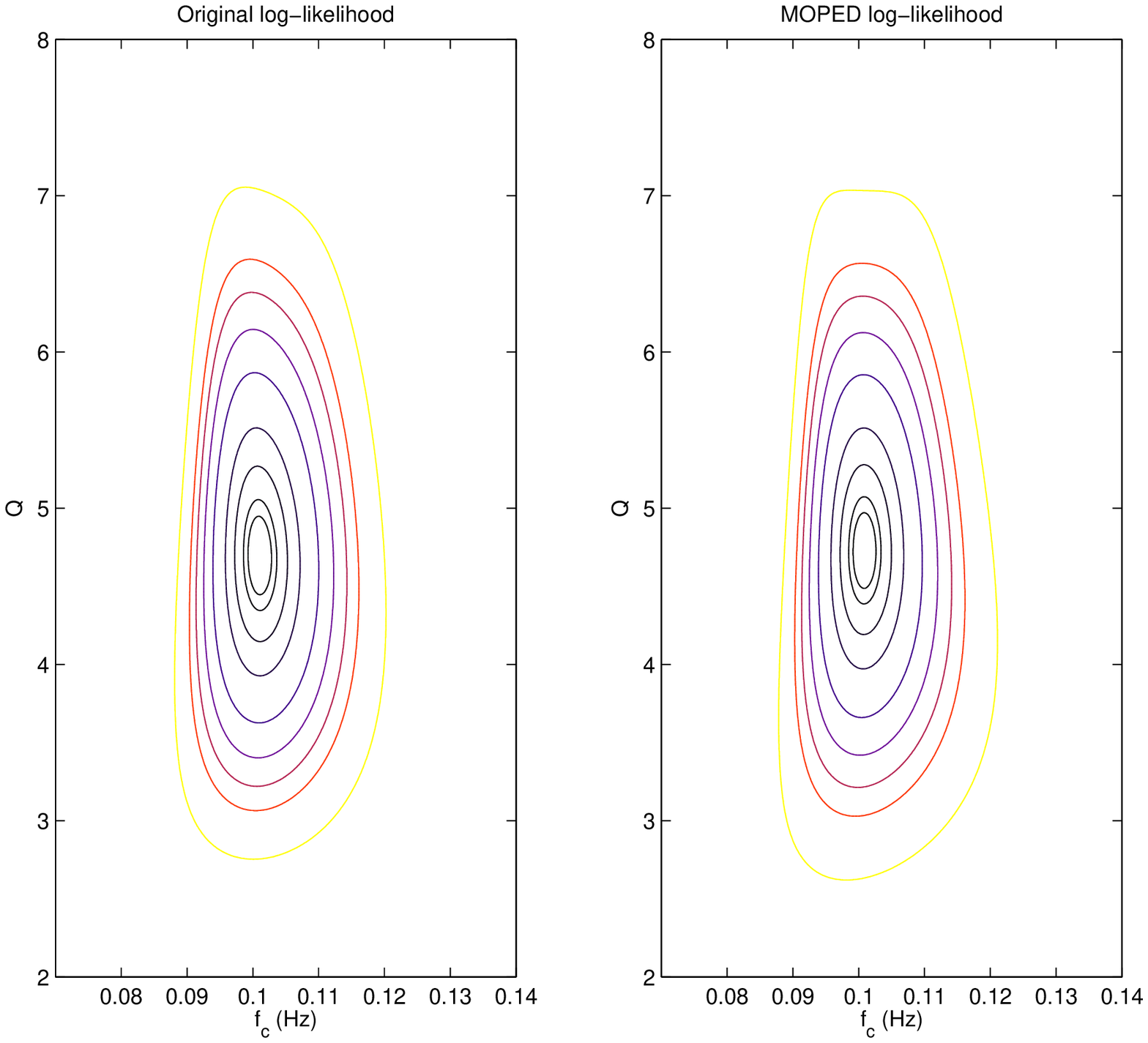}
\caption{Contours of the original and MOPED log-likelihoods (left and right, respectively).  The MOPED likelihood has been multiplied by a constant factor so that its peak value is equal to the peak of the original likelihood.  Contours are at 1, 2, 5, 10, 20, 30, 40, 50, 75, and 100 log-units below the peak going from the inside to outside.}
\label{fig:FQlikes}
\end{center}
\end{figure}

Interestingly, we find that even when MOPED fails in a one-parameter case, adding a second parameter may actually rectify the problem, although not necessarily.  If we now allow the width of the burst, $Q$, to be a variable parameter, there are now two orthognal MOPED weighting vectors that need to be calculated.  This gives us two compressed parameters for each pair of $f_c$ and $Q$.  Each of these may have its own unphysical degeneracies, but in order to give an unphysical mode of equal likelihood to the true peak, these degeneracies will need to coincide.  In Figure~\ref{fig:YtrueContours}, we plot the contours in $(f_c,Q)$ space of where $\left<y_i\right>(\btheta;\btheta_F) = \left<y_i\right>(\hat{\btheta}_M;\btheta_F)$ as $\btheta$ ranges over $f_c$ and $Q$ values.  We can clearly see the degeneracies present in either variable, but since these only overlap at the one location, near to where the true peak is, there is no unphysical second mode in the MOPED likelihood.
Hence, when we plot the original and MOPED log-likelihoods in Figure~\ref{fig:FQlikes}, although the behaviour away from the peak has changed, the peak itself remains in the same location and there is no second mode.

\begin{figure}
\begin{center}
\includegraphics[width=3.25in]{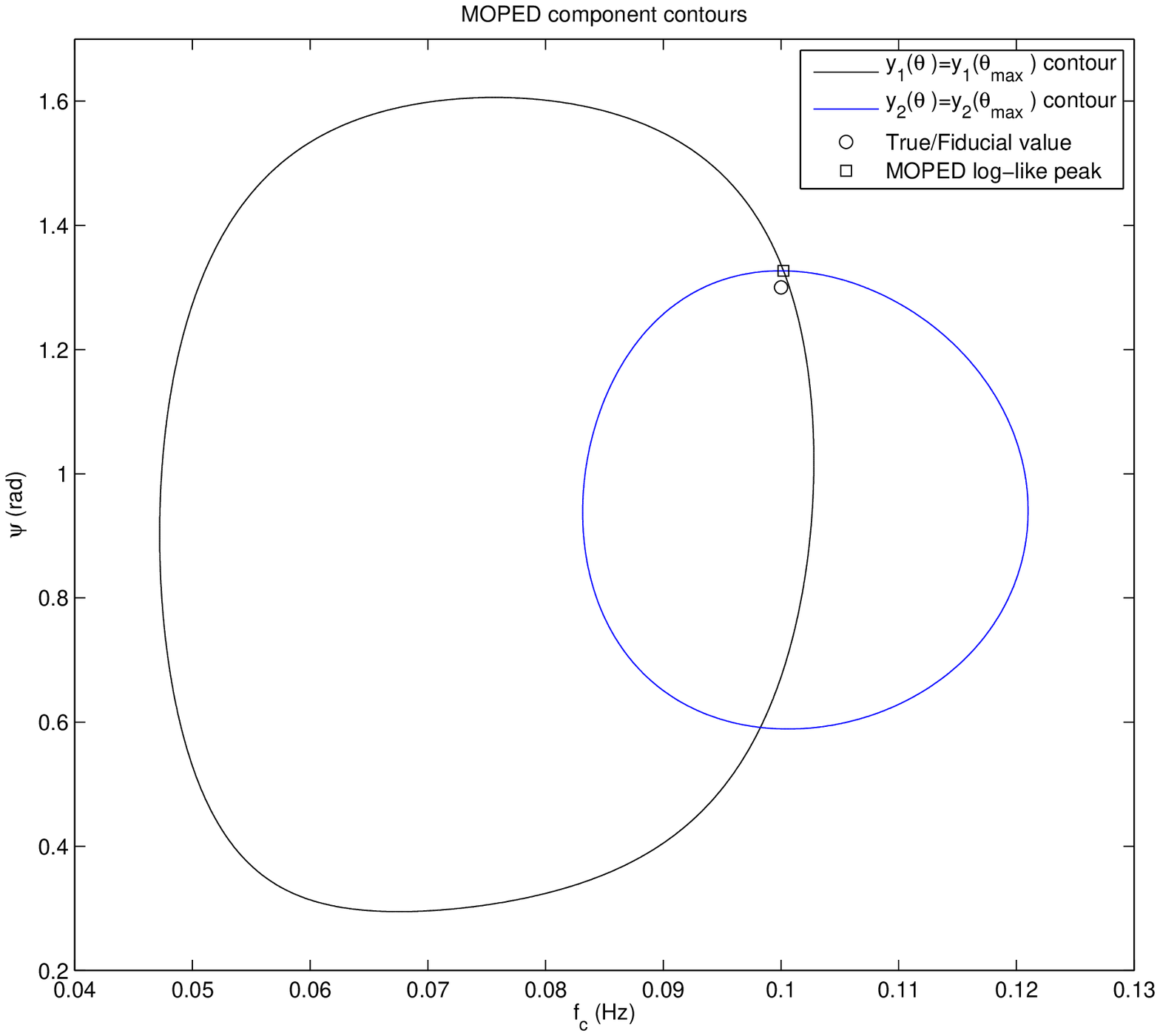}
\caption{Contours of $\left<y_1\right>(\btheta;\btheta_F) = \left<y_1\right>(\hat{\btheta};\btheta_F)$ and $\left<y_2\right>(\btheta;\btheta_F) = \left<y_2\right>(\hat{\btheta};\btheta_F)$ values as they vary as functions of $f_c$ and $\psi$.}
\label{fig:YtrueContours3}
\end{center}
\end{figure}

\begin{figure}
\begin{center}
\includegraphics[width=3.25in]{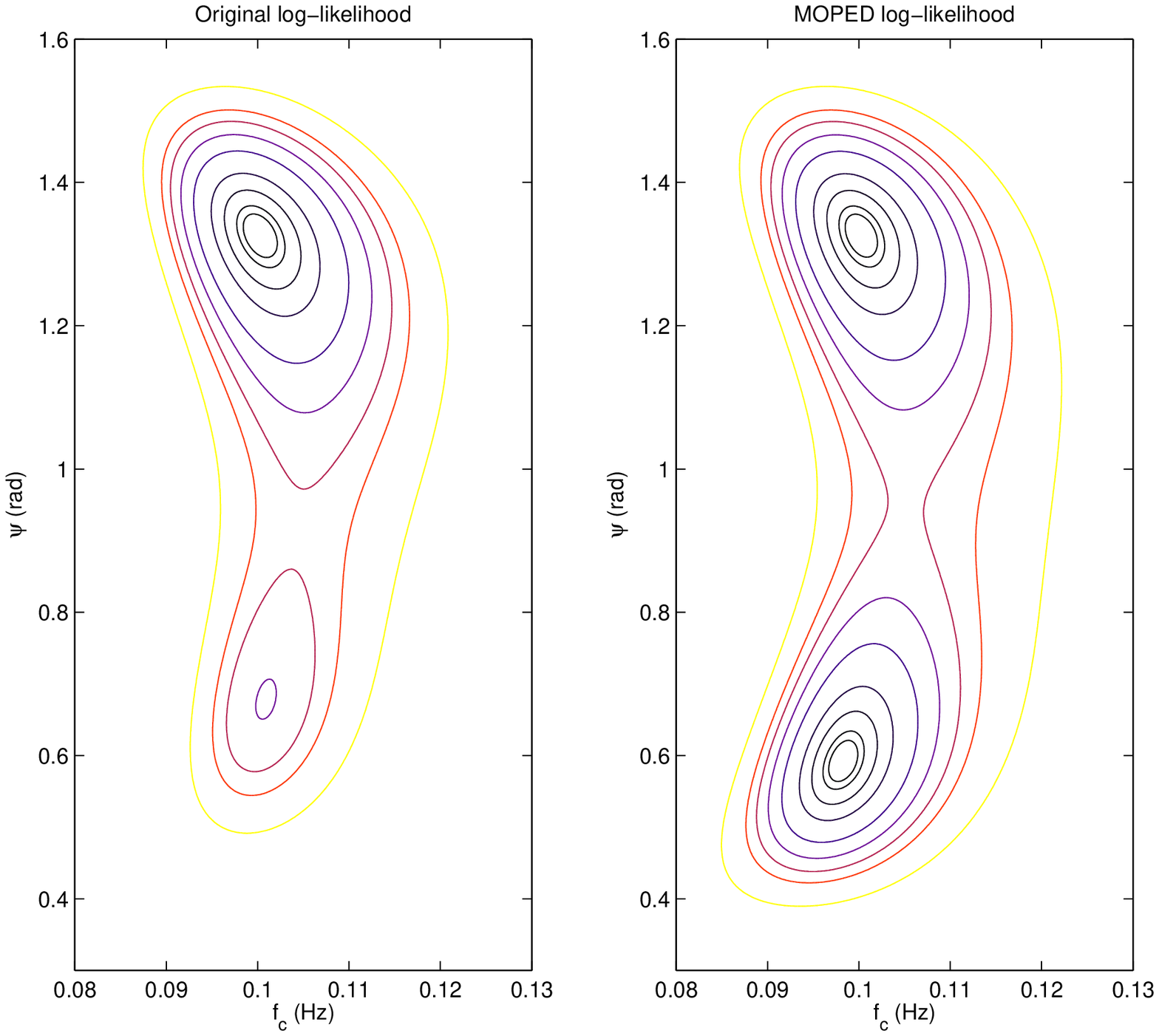}
\caption{Contours of the original and MOPED log-likelihoods (left and right, respectively).  The MOPED likelihood has been multiplied by a constant factor so that its peak value is equal to the peak of the original likelihood.  Contours are at 1, 2, 5, 10, 20, 30, 40, 50, 75, and 100 log-units below the peak going from the inside to outside.}
\label{fig:FPslikes}
\end{center}
\end{figure}

Adding more parameters, however, does not always improve the situation.  We now consider the case where $Q$ is once again fixed to its true value and we instead make the polarisation of the burst, $\psi$, a variable parameter.  There are degeneracies in both of these parameters and in Figure~\ref{fig:YtrueContours3} we plot the contours in $(f_c,\psi)$-space where the compressed values are each equal to the value at the maximum MOPED likelihood point.  These two will necessarily intersect at the maximum likelihood solution, near the true value ($f_c=0.1$ Hz and $\psi=1.3$ rad), but a second intersection is also apparent.  This second intersection will have the same likelihood as the maximum and be another mode of the solution.  However, as we can see in Figure~\ref{fig:FPslikes} in the left plot, this is not a mode of the original likelihood function.  MOPED has, in this case, created a second mode of equal likelihood to the true peak.

\begin{figure}
\begin{center}
\includegraphics[width=3.25in]{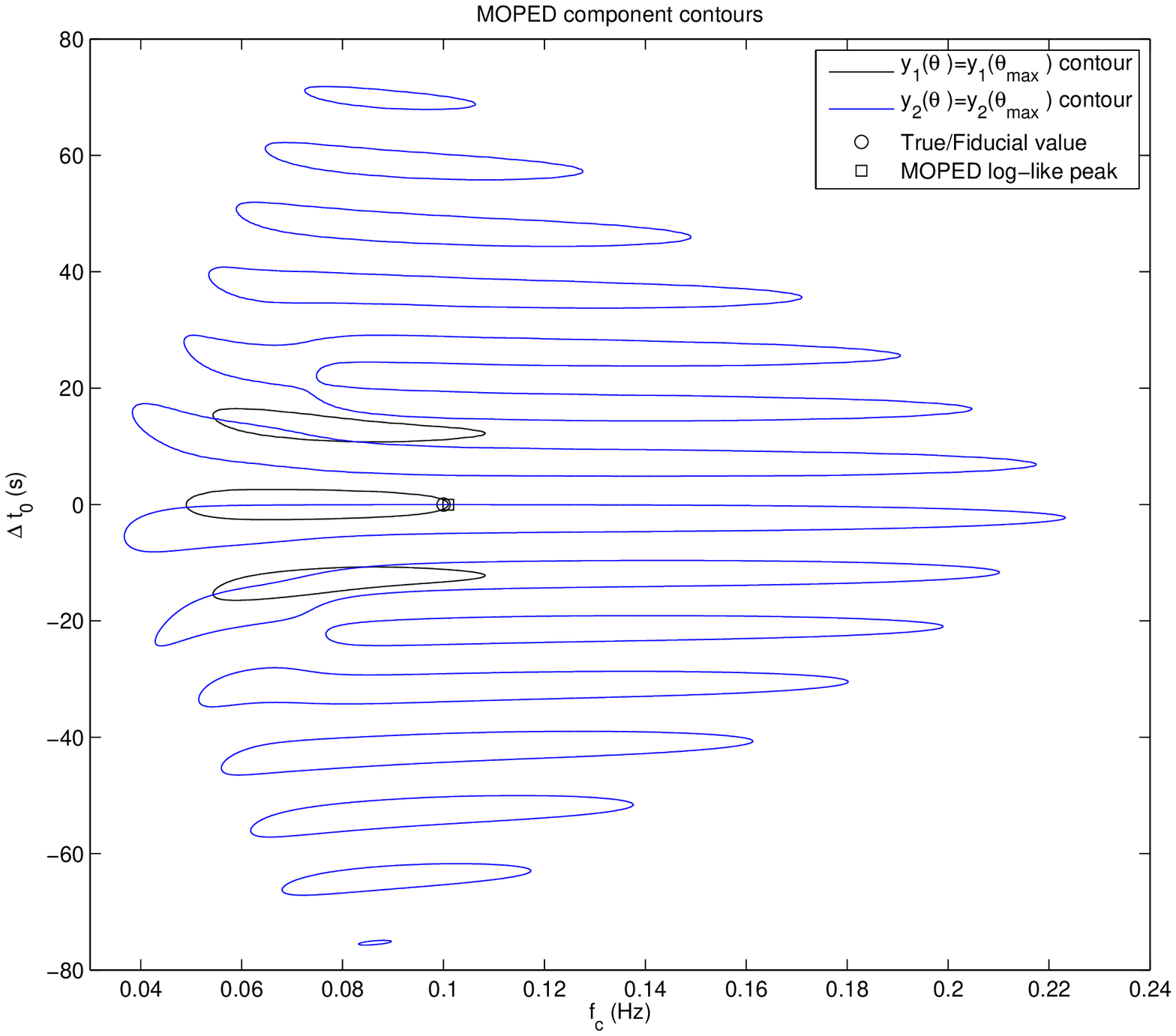}
\caption{Contours of $\left<y_1\right>(\btheta;\btheta_F) = \left<y_1\right>(\hat{\btheta};\btheta_F)$ and $\left<y_2\right>(\btheta;\btheta_F) = \left<y_2\right>(\hat{\btheta};\btheta_F)$ values as they vary as functions of $f_c$ and $t_0$.  We can see many intersections here other than the true peak.}
\label{fig:YtrueContours2}
\end{center}
\end{figure}

\begin{figure}
\begin{center}
\includegraphics[width=3.25in]{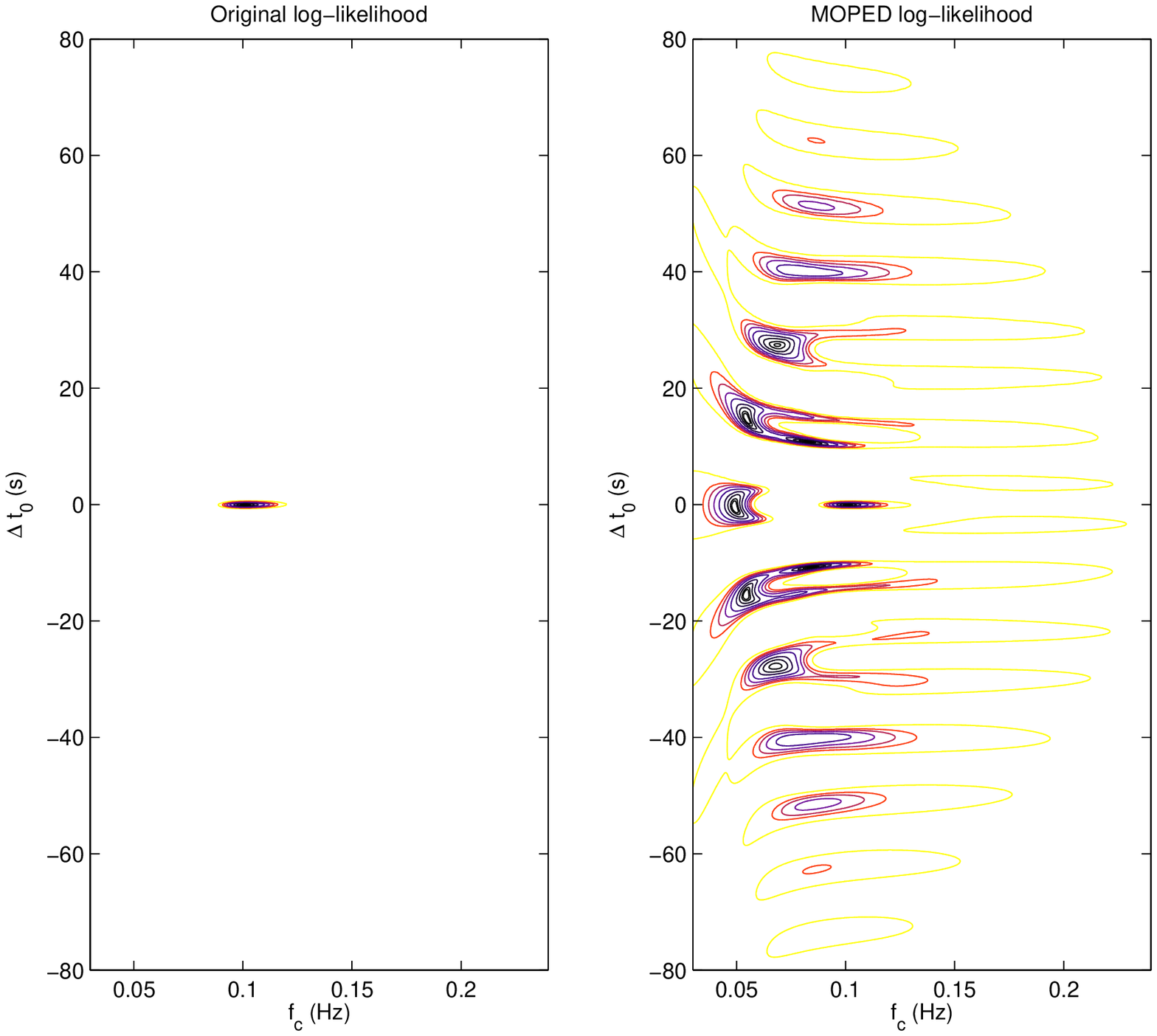}
\caption{Contours of the original and MOPED log-likelihoods (left and right, respectively).  The MOPED likelihood has been multiplied by a constant factor so that its peak value is equal to the peak of the original likelihood.  Contours are at 1, 2, 5, 10, 20, 30, 40, 50, 75, and 100 log-units below the peak going from the inside to outside.}
\label{fig:FTlikes}
\end{center}
\end{figure}

For an even more extreme scenario, we now fix to the true $\psi$ and allow the time of arrival of the burst $t_0$ to vary (we also define $\Delta t_0 = t_0-t_{0,T}$).  In this scenario, the contours in $(f_c,\Delta t_0)$-space where $\left<y_i\right>(\btheta;\btheta_F) = \left<y_i\right>(\hat{\btheta}_M;\btheta_F)$ are much more complicated.  Thus, we have many more intersections of the two contours than just at the likelihood peak near the true values and MOPED creates many alternative modes of likelihood equal to that of the original peak. This is very problematic for parameter estimation.  In Figure~\ref{fig:YtrueContours2} we plot these contours so the multiple intersections are apparent.  Figure~\ref{fig:FTlikes} shows the original and MOPED log-likelihoods, where we can see the single peak becoming a set of peaks.

\section{Discussion and Conclusions}
What we can determine from the previous two sections is a general rule for when MOPED will generate additional peaks in the likelihood function equal in magnitude to the true one.  For an $M$-dimensional model, if we consider the $(M-1)$-dimensional hyper-surfaces where $\left<y_i\right>(\btheta;\btheta_F) = \left<y_i\right>(\hat{\btheta}_M;\btheta_F)$, then any point where these $M$ hyper-surfaces intersect will yield a set of $\btheta$-parameter values with likelihood equal to that at the peak near the true values.  We expect that there will be at least one intersection at the likelihood peak corresponding to approximately the true solution.  However, we have shown that other peaks can exist as well.  The set of intersections of contour surfaces will determine where these additional peaks are located.  This degeneracy will interact with the model's intrinsic degeneracy, as any degenerate parameters will yield the same compressed values for different original parameter values.

Unfortunately, there is no simple way to find these contours other than by mapping out the $\left<y_i\right>(\btheta;\btheta_F)$ values, which is equivalent in procedure to mapping the MOPED likelihood surface.  The benefit comes when this procedure is significantly faster than mapping the original likelihood surface.  The mapping of $\left<y_i\right>(\btheta;\btheta_F)$ can even be performed before data is obtained or used, if the fiducial model is chosen in advance; this allows us to analyse properties of the MOPED compression before applying it to data analysis.  If the MOPED likelihood is mapped and there is only one contour intersection, then we can rely on the MOPED algorithm and will have saved a considerable amount of time, since MOPED has been demonstrated to provide speed-ups of a factor of up to $10^7$ in~\cite{MOPED2}.  However, if there are multiple intersections then it is necessary to map the original likelihood to know if they are due to degeneracy in the model or were created erroneously by MOPED.  In this latter case, the time spent finding the MOPED likelihood surface can be much less than that which will be needed to map the original likelihood, so relatively little time will have been wasted.  If any model degeneracies are known in advance, then we can expect to see them in the MOPED likelihood and will not need to find the original likelihood on their account.

One possible way of determining the validity of degenerate peaks in the MOPED likelihood function is to compare the original likelihoods of the peak parameter values with each other.  By using the maximum MOPED likelihood point found in each mode and evaluating the original likelihood at this point, we can determine which one is correct.  The correct peak and any degeneracy in the original likelihood function will yield similar values to one another, but a false peak in the MOPED likelihood will have a much lower value in the original likelihood and can be ruled out.  This means that a Bayesian evidence calculation cannot be obtained from using the MOPED likelihood; however, the algorithm was not designed to be able to provide this.

The solution for this problem presented in~\cite{Protopapas} is to use multiple fiducial models to create multiple sets of weighting vectors. The log-likelihood is then averaged across these choices. Each different fiducial will create a set of likelihood peaks that include the true peaks and any extraneous ones. However, the only peaks that will be consistent between fiducials are the correct ones. Therefore, the averaging maintains the true peaks but decreases the likelihood at incorrect values. This was tested with 20 random fiducials for the two-parameter models presented and was found to leave only the true peak at the maximum likelihood value. Other, incorrect, peaks are still present, but at log-likelihood values five or more units below the true peak. When applied to the full seven parameter model, however, the SNR threshold for signal recovery is increased significantly, from $\simeq 10$ to $\simeq 30$.

The MOPED algorithm for reducing the computational expense of likelihood functions can, in some examples, be extremely useful and provide orders of magnitude of improvement.  However, as we have shown, this is not always the case and MOPED can produce erroneous peaks in the likelihood that impede parameter estimation.  It is important to identify whether or not MOPED has accurately portrayed the likelihood function before using the results it provides. Some solutions to this problem have been presented and tested, 

\section*{Acknowledgments}
PG's PhD is funded by the Gates Cambridge Trust.

\label{lastpage}

\end{document}